\documentclass[runningheads]{llncs}
\usepackage[T1]{fontenc}
\usepackage[hidelinks]{hyperref}

\usepackage{graphicx}
\usepackage{booktabs}
\usepackage{multirow}
\usepackage{array}
\usepackage[linesnumbered,ruled,vlined]{algorithm2e}
\usepackage{amsmath}
\usepackage{enumitem} 
\usepackage{float}
\usepackage[
  subtle
]{savetrees}
\usepackage{blindtext}

\setlist{itemsep=2pt, topsep=2pt}

\begin{document}
\title{Skelite: Compact Neural Networks for Efficient Iterative Skeletonization}
%
%
\author{Luis D. Reyes Vargas \inst{1} \and Martin J. Menten \inst{2,3,4} \and Johannes C. Paetzold \inst{5} \and Nassir Navab \inst{1,4} \and Mohammad Farid Azampour \inst{1,4}}
\authorrunning{Anonymous}
%
\institute{Computer Aided Medical Procedures, Technical University of Munich, Germany \and Chair for AI in Healthcare and Medicine, Technical University of Munich, Germany \and BioMedIA, Department of Computing, Imperial College London, UK \and Munich Center for Machine Learning (MCML), Germany \and Weill Cornell Medicine, Cornell University, New York City, NY, USA}
\maketitle              
\begin{abstract}

Skeletonization extracts thin representations from images that compactly encode their geometry and topology. These representations have become an important topological prior for preserving connectivity in curvilinear structures, aiding medical tasks like vessel segmentation. Existing compatible skeletonization algorithms face significant trade-offs: morphology-based approaches are computationally efficient but prone to frequent breakages, while topology-preserving methods require substantial computational resources. 

We propose a novel framework for training iterative skeletonization algorithms with a learnable component. The framework leverages synthetic data, task-specific augmentation, and a model distillation strategy to learn compact neural networks that produce thin, connected skeletons with a fully differentiable iterative algorithm. 


Our method demonstrates a $100\times$ speedup over topology-constrained algorithms while maintaining high accuracy and generalizing effectively to new domains without fine-tuning. Benchmarking and downstream validation in 2D and 3D tasks demonstrate its computational efficiency and real-world applicability\footnotetext[1]{Code and data available at: \href{https://github.com/luisdavid64/Skelite}{\texttt{https://github.com/luisdavid64/Skelite}}}.

\end{abstract}
\section{Introduction}

Skeletonization algorithms reduce the foreground of an image to a skeletal representation that approximates the medial axis, defined as the set of points with more than one closest point on the shape's boundary \cite{blum_grassfire}. These representations condense the topological and geometric properties of an object, making them useful in a variety of medical applications such as blood flow analysis, image registration, and surgical planning \cite{Lidayov2017SkeletonbasedFF,image_registration,Fridman2004ExtractingBT}. Skeleton extraction has been widely explored for digital images with the core principles of accuracy and computational efficiency, leading to a vast body of works \cite{bloomberg_thickening,Lobregt_skel,couprie_asymmetric_simple_skel,palagyi_sequential_simple_skel,palagyi_subiteration_simple_skel,bertrand_bool_char,zhang_simple_skel}.

\begin{figure}[htbp] 
    \centering
\includegraphics[width=\textwidth]{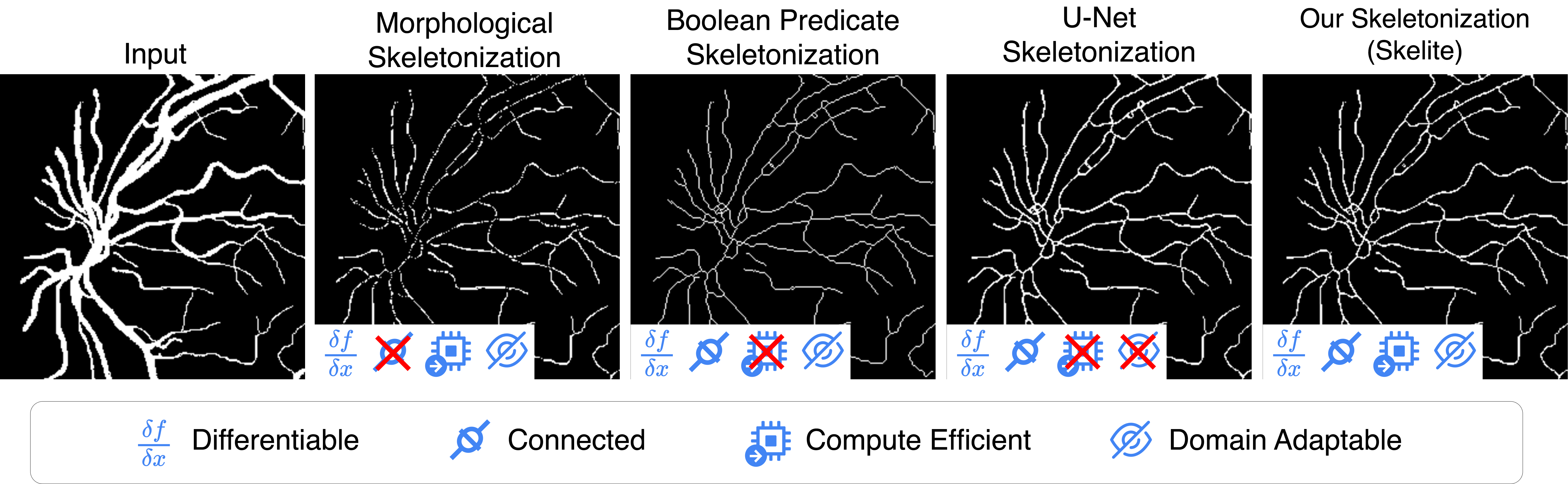}
    \caption{Existing skeletonization methods trade off speed, accuracy, and adaptability. Morphology-based methods often break skeletons, Boolean predicates are slow but topologically accurate, and U-Nets struggle with domain shifts. Our method is fast, adaptable, and produces thin skeletons with few breakages.}
    \label{fig:abstract} 
\end{figure}


Recently, skeletonization has been adopted as a structural prior in deep learning for semantic segmentation. Studies have demonstrated that loss functions defined using a differentiable skeletonization algorithm result in better preservation of tubular structures~\cite{dynamic_skel_edge_detection,shit_diff_skel,clCe}. 


Unfortunately, most established skeletonization methods cannot be trivially adapted to be differentiable, reducing the available algorithms to a select few based on iterative boundary thinning. Inspired by morphological operations, Shit \textit{et al}. introduced a fast, differentiable skeletonization algorithm based on iterative applications of erosion and dilation operations. \cite{shit_diff_skel}. However, this approach, which we refer to as the Morphological method, frequently produces breakages in skeletons. In response, a skeletonization algorithm, which we refer to as the Boolean method, was developed that guarantees topological correctness using Boolean predicate kernels for simple point detection at a higher computational cost~\cite{menten_skel}.

The advances in convolutional neural networks (CNNs) have also produced a class of learned approaches that frame skeletonization as a segmentation problem \cite{nguyen_unet_skel,panichev_unet_skel}. These methods have demonstrated promise in domain-specific challenges \cite{Skelneton} supported by advances in semantic segmentation, often leveraging the U-Net architecture \cite{unet}. While inherently compatible with backpropagation, these solutions are susceptible to domain shifts and impose resolution constraints on the inputs, complicating their use in downstream tasks where geometric conditions vary. 

Neural network solutions follow the trend of increasing parameters for improved performance. However, traditional iterative skeletonization algorithms typically rely on a small set of predefined neighborhood patterns, applied repeatedly, to achieve accurate skeletonization \cite{zhang_simple_skel,bertrand_bool_char,Lee1994BuildingSM}.

Based on this intuition, we propose a novel approach for skeletonization that combines iterative skeletonization and neural networks. This scheme, which we call \textbf{\textit{Skelite}}, requiring only a few learnable convolutional filters coupled with ReLU activations, max-pooling, and matrix additions, which makes it fully compatible with gradient-based optimization.


Our proposed method needs training only on synthetic data, showcasing robust generalization in curvilinear tasks across diverse applications without requiring domain-specific fine-tuning. Unlike other neural network algorithms, our approach allows access to intermediate states of the skeletonization, making the process more transparent and explainable. Our experiments demonstrate that Skelite captures features that improve the connectivity of the resulting skeletons compared to the morphological approach.
We evaluate the proposed algorithm's efficacy in the downstream task of segmentation and observe improved performance with minimal computational overhead. This efficiency makes our algorithm more accessible than other methods explicitly addressing topological errors.

To summarize, the contributions of our work include: 
\begin{itemize}[label=\textbullet]
    \item An iterative skeletonization algorithm with a learnable component that is compatible with gradient-based optimization.
    \item A training framework for the proposed approach, including a synthetic training dataset, task-specific augmentation, and a model distillation strategy.
    \item Extensive experiments on 2D and 3D datasets, demonstrating computational efficiency and generalization to unseen curvilinear datasets for generating thin skeletons (See Figure \ref{fig:abstract}).
    \item Showcasing improvements on four datasets in preserving continuity when integrated into a segmentation pipeline.
\end{itemize}

\section{Method}


Given an image $I$ and a target skeleton $S$, we aim to find a function $f_\theta$, parameterized with a convolutional neural network, that approximates the skeleton in $N$ steps.

\begin{equation}
    \begin{aligned}
        S^{0} &= I \\
        S^{n+1} &= f_\theta(S^n) \\
        S &\approx S^N
    \end{aligned}
\end{equation}

Following this scheme, Algorithm \ref{alg:skelite} describes our skeletonization method, where the operator $\odot$ denotes element-wise multiplication. On a high level, each iteration consists of three steps:

\begin{enumerate}
    \item \textbf{Deletion proposal:} We obtain a proposal of points for deletion through morphological boundary extraction. An eroded input is subtracted from the original input to obtain the boundary. 
    \item  \textbf{Proposal Evaluation}: The network is fed the original input, boundary and current skeleton to evaluate which points can be deleted, producing a delta mask.
    \item \textbf{Skeleton and Image Update}: The skeleton is updated by subtracting the delta mask, and the input is updated with the eroded input for the next iteration.
\end{enumerate}

Erosion is performed using max-pooling, as described in \cite{shit_diff_skel}, ensuring that all operations remain differentiable. At each iteration, the CNN receives two types of inputs: the boundary of the image as the primary deletion target, and two contextual images to guide the network's predictions. The first contextual input is the eroded image, which reveals points not currently considered for deletion to encourage the network to remove points in a way that avoids future discontinuities. The second input is the current skeletal representation, helping the network avoid deleting points that would disrupt continuity.

The method discussed is designed for binary images; however, in many practical applications, the inputs can include matrices of continuous values. To handle these continuous inputs, we adopt the stochastic discretization technique described by Menten \textit{et al.} \cite{menten_skel}, denoted as `binarize' in line \ref{line:1} of Algorithm \ref{alg:skelite}. This technique enables binarization while preserving gradient flow through the operation with a straight-through estimator, facilitating optimization despite the discrete nature of binarization.

\begin{algorithm}[t] 
\caption{Skeletonization Procedure.
$f_\theta$ is our skeletonization network, $I$ is the input image, and $N$ is the number of iterations.} 
\label{alg:skelite}
\KwIn{$f_\theta$, $I$, $N$ }
$I \gets \text{binarize}(I)$ \label{line:1}\\
$S \gets I$ \\
\For{$i \gets 1$ \text{to} $N$}{
    $eroded \gets$ \text{erode}(\text{erode}(I)) \\
    $boundary \gets I - eroded$ \\
    $delta \gets$ $f_\theta(I, boundary, S)$ \\
    $delta \gets delta \odot boundary$ \\
    $S \gets S - delta$ \\
    $I \gets eroded$
}
\KwOut{$S$}
\end{algorithm}

\begin{figure}[t]
    \centering
 \includegraphics[width=0.6\textwidth]{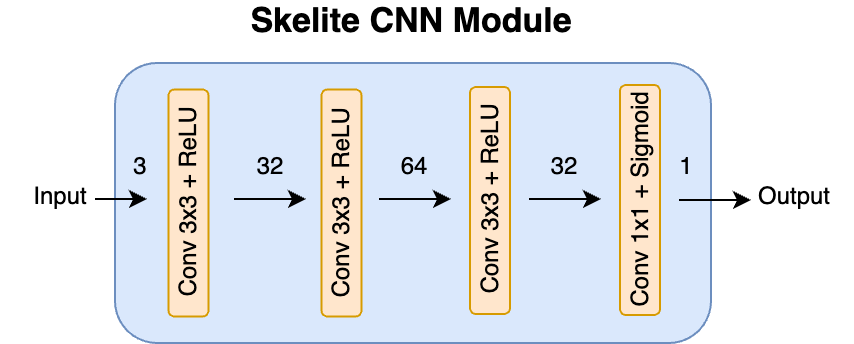}
    \caption{Our neural network skeletonization module. The arrows indicate the number of output channels after each layer.}
    \label{fig:arch-net}
\end{figure}

Figure \ref{fig:arch-net} illustrates the architecture of our lightweight convolutional network. It consists of
$3\times 3$ convolutional layers, ReLU activations, and a sigmoid activation after the last layer. It does not include downsampling and uses stride-1 convolutions throughout, allowing it to work with inputs of any resolution.

\subsection{Model Supervision and Compression}

\begin{figure}
    \centering
    \includegraphics[width=\textwidth]{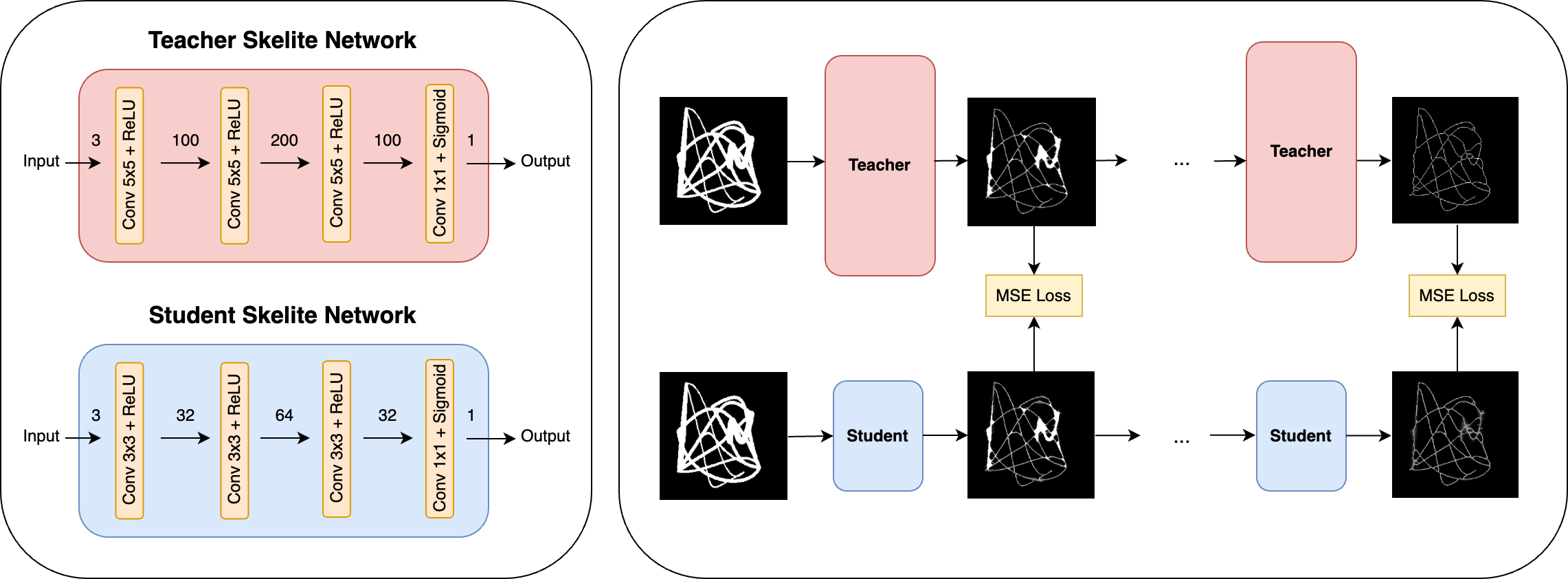}
    \caption{Example of knowledge distillation strategy. The student network can receive very tight supervision at each skeletonization step from the teacher network.}
    \label{fig:compression-strat}
\end{figure}

Skeletal data is inherently sparse, with skeletal points constituting less than 5\% of the total in our experimental data. This makes the task of classifying skeletal points highly imbalanced. Previous works on learnable skeletonization adopt a combination of the focal loss \cite{focal_loss} and Dice loss to address the issue \cite{panichev_unet_skel,nguyen_unet_skel}.  However, we find these losses can lead to insufficient thinning in the skeletons. 

To address this, we introduce a \textbf{neighborhood loss} $L_{neighborhood}$. This loss uses an all ones $m \times m$ (or $m\times m \times m$ in 3D) convolutional kernel K. Given a predicted skeleton $S_P$ and ground truth skeleton $S_T$, we apply the kernel to obtain the representations $S_{P_K} = S_P * K$ and $S_{T_K} = S_K * K$. We utilize the mean absolute error to obtain $L_{neighborhood} = \text{MAE}(S_{P_K}, S_{T_K})$.

The loss is empirically designed to utilize the sparse neighborhoods of skeletons, encouraging thinner predictions by ensuring similarity between the local neighborhoods of the predicted and ground truth skeletons. To avoid over-thinning, we integrate the focal loss $L_{focal}$ and the Dice loss $L_{Dice}$ to account for overlap to obtain the loss on the final outputs of our algorithm.

\begin{equation}
    L = L_{focal} + L_{Dice} + L_{neighborhood}
\end{equation}

In our experiments, we observe that large Skelite models are more effective at learning skeletonization from data. We aim to retain this performance in smaller, faster models. Drawing inspiration from knowledge distillation \cite{Bucila2006ModelC,Hinton2015DistillingTK}, we propose a method to transfer the performance of a Skelite model into a more compact version. As shown in Figure \ref{fig:compression-strat}, our distillation strategy takes advantage of the iterative nature of the skeletonization process, providing supervision at each step. This fine-grained supervision enables the smaller model to effectively learn skeletonization by leveraging the guidance from each stage of the process of a similarly structured algorithm.




\subsection{Synthetic Data: Bézier Dataset} \label{sec:bezier_data}

\begin{figure}
    \centering
    \includegraphics[width=0.9\linewidth]{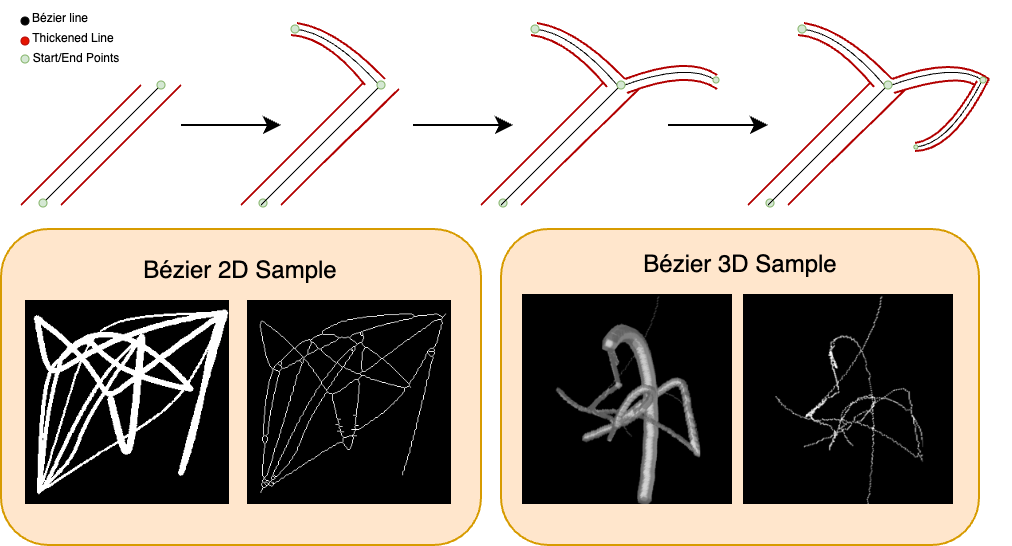}
    \caption{Top: Illustration of Bézier dataset generation. Bottom: representative samples in 2D and 3D.}
    \label{fig:generation}
\end{figure}    

We wish to learn skeletonization from data that mimics the appearance and characteristics of vasculature or other curvilinear structures. To this end, we propose a procedural method to generate synthetic data using cubic Bézier curves, described by the formula:

\begin{equation}
    B(t) = (1 - t)^3 P_0 + 3(1 - t)^2 t P_1 + 3(1 - t) t^2 P_2 + t^3 P_3,\  t \in [0,1]
\end{equation}

where $P_0$ is the starting point of the curve, $P_3$ is the end point, $P_1$ and $P_2$ are control points and $t$ is the interpolation parameter. The process is detailed in figure \ref{fig:generation}.

The samples are initialized with a set of primary Bézier curves, referred to as trunks, whose control points are randomly sampled within the image boundaries. The thickness of these trunks is uniformly sampled from the integer range $[w_{trunk}, W_{trunk}]$. From the endpoints of these trunks, a branching process begins, in which a number of Bézier curves are sampled to grow from this point in the range $[n_{trunk}, N_{trunk}]$. The thickness of these branching curves is sampled similarly to the trunks but is constrained such that their thickness cannot exceed that of their parent curve. To simulate hierarchical growth, the process is repeated recursively, with a pre-defined recursion depth serving as the stopping condition for the generation process. We obtain the skeleton labels for the training data by leveraging the Boolean skeletonization algorithm for its topological correctness.

The generation process approximates patterns commonly observed in curvilinear structures, such as branching and smooth curved geometries. We use the same synthetic data generation strategy in 2D and 3D, and generate 5000 samples for 2D and 20000 samples for 3D which we will make publicly available. To further increase the diversity of the data, we introduce a complementary augmentation.

\subsection{Thickening Augmentation}

Complementary to the diversity of the data generation process, we produce a thickening augmentation to enhance the geometric diversity of samples during training. Thickening can be derived as a byproduct of background thinning \cite{bloomberg_thickening}. However, in our approach we leverage the availability of paired reference images and skeletons for a straightforward process that ensures topology preservation. Specifically, given a reference image $I$ and its corresponding skeleton $S$, we can obtain intermediate thickened representations $S_{thick}$ through dilation. The process is outlined by the following equation.

\begin{equation}
S_{\text{thick}} = \text{dilate}^k(S) \odot I
\end{equation}

Where we let $\text{dilate}^k(S)$ represent the result of applying dilation $k$ times to the skeleton. Thus, the thickening is controlled by the number of dilation applications, which is uniformly sampled from the integers between zero and the maximum thickness from the generated samples. Here, the multiplication with the reference image $I$ ensures that the thickened skeleton remains consistent with the original topology and geometry of the object. When $k=0$, the augmentation returns the skeleton as is, presenting the challenge of skeleton preservation.

\section{Experiments and Results}



We validate our skeletonization method in three ways. First, we explore skeletonization results in the synthetic data domain and evaluate its transition to real-world data compared to other neural network methods. Next, we assess its topological correctness, geometry, computational efficiency, and overlap with a baseline skeleton obtained using algorithms by Zhang et al. (2D) \cite{zhang_simple_skel} and Lee et al. (3D) \cite{Lee1994BuildingSM}, as implemented in the widely-used scikit-image library \cite{scikit}. Finally, we demonstrate its practical applicability by integrating it with clDice for deep learning-based binary segmentation.

\subsection{Datasets}

Our experiment data comprises a synthetic dataset and four publicly available datasets containing thin structures.

\begin{itemize}[label=\textbullet]

\item Bézier consists of $5000$ samples in 2D and $20000$ samples in 3D. For these experiments, we set the parameters from section \ref{sec:bezier_data} to  $w_{trunk} = 3$, $W_{trunk} = 10$, $n_{branches} = 1$, $N_{branches} = 3$.

\item DRIVE \cite{Drive_Dataset} (Digital Retinal Images for Vessel Extraction) consists of $40$ 2D color fundus images annotated with the visible retinal vessels.

\item ROADS \cite{roads_dataset} (Massachussets Roads) is dataset for studying the segmentation of roads from aerial images. It consists of 1171 images with annotations. 

\item ASOCA \cite{Asoca} (Automated Segmentation of Coronary Arteries) consists of $40$ 3D cardiac CT angiography scans depicting $20$ normal and $20$ diseased coronary arteries.

\item TopCoW \cite{TopCow} (Topology-Aware Anatomical Segmentation of the Circle of Willis) is a 3D vessel segmentation challenge containing $250$ samples of CTA and MRA data depicting the Circle of Willis vessels.
\end{itemize}

\subsection{Evaluation Metrics}

We use a diverse set of metrics to benchmark our method and evaluate performance in segmentation, including overlap, connectivity, and topological accuracy. We include the Dice similarity coefficient as our measure of overlap and supplement it with the clDice metric, which was introduced as a complement to the loss function to evaluate connectivity in tubular structure segmentation \cite{shit_diff_skel}. As in previous works \cite{menten_skel}, we assess the topological correctness of our method by evaluating the Betti number errors $\beta_0$ and $\beta_1$.


Inspired by previous works on skeleton evaluation, we extend our analysis with proxy metrics to assess the thickness of skeletons \cite{systematic_evaluation_skel,coral_metrics,similarity_metric_skel}. Namely, we derive two metrics from the Euclidean distance transform. The Euclidean distance transform generates an image representation where each foreground pixel is replaced by the shortest Euclidean distance to the background. By taking the maximum (99\textsuperscript{th} percentile) and average of these distances, we obtain two metrics: maximum thickness, which expresses the largest distance, and average thickness, which represents the average distance from the foreground to the background of the image. These metrics approximately encode the thickness of the patterns, so we name them `99\textsuperscript{th} max thickness' and `average thickness' for intuitive appeal.

\subsection{Transitioning from Synthetic to Real world data}

\begin{figure}[t]
    \centering
    \includegraphics[width=\textwidth]{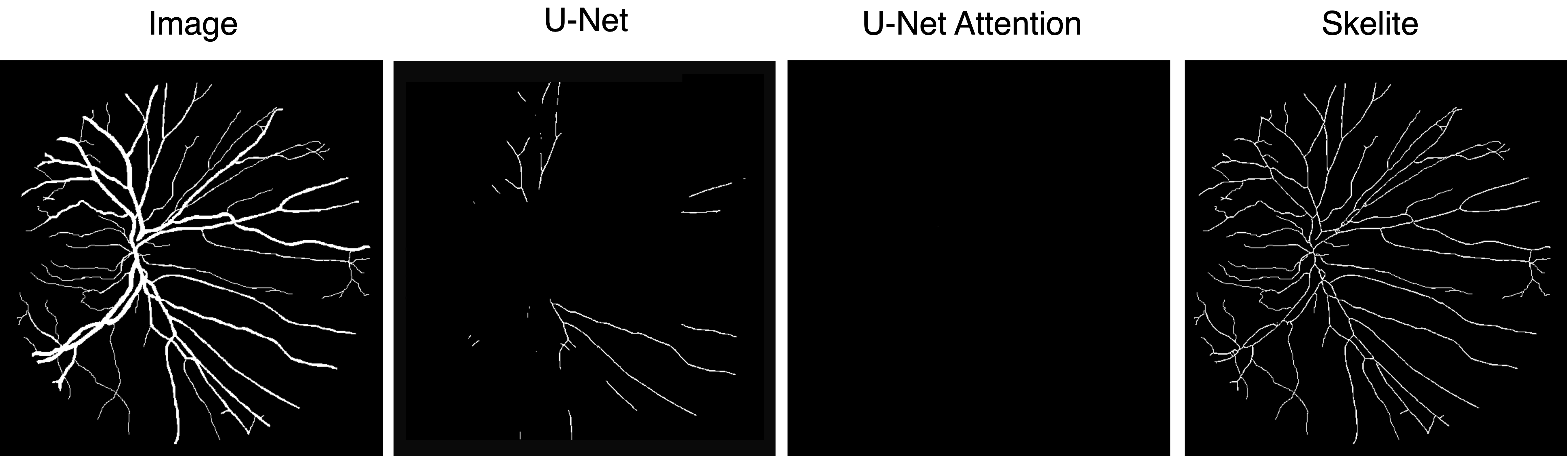}
    \caption{A Comparison of neural network skeletonization trained on the Bézier dataset and tested on DRIVE. U-Net-based methods overfit to Bézier-specific features, resulting in over-thinning in DRIVE. Skelite's controlled iterative thinning addresses this issue, allowing it to generalize to new domains more easily.}
    \label{fig:drive-networks}
\end{figure}

In this experiment, we evaluate the performance of our method on the synthetic dataset and assess its generalizability to unseen real-world data after training on the Bézier dataset. We compare our approach with two other neural network skeletonization archxitectures, all trained using the same dataset and augmentation strategy, and then tested on DRIVE:

\begin{itemize}[label=\textbullet]

    \item U-Net: Proposed by Panichev \textit{et al.} \cite{panichev_unet_skel}, this method leverages a U-Net architecture for skeleton extraction.
    
    \item U-Net Attention: Introduced by Nguyen \cite{nguyen_unet_skel}, this approach includes the attention mechanism in the U-Net for feature extraction \cite{nguyen_cbam}. 
\end{itemize}

\begin{table}[b]
\centering
\caption{We test the generalization of neural network methods for skeleton extraction by training on synthetic data and testing on real-world data.}
\resizebox{\textwidth}{!}{%
\setlength{\tabcolsep}{8pt} 
\begin{tabular}{@{}llcccccc@{}}
\toprule
Dataset & Method & $\beta_0$ error & $\beta_1$ error & Avg. Thickness & 99\textsuperscript{th} Max Thickness  & Dice & Run time [ms] \\
\midrule
\multirow{3}{*}{\textbf{2D Bézier}} 
& U-Net & 0.07 & 2.92 & 1.07 & 3.77 & 0.72 & 43 \\
& U-Net Attention & 0.02 & 4.28 & 1.03 & 2.35 & 0.75 & 31 \\
& Skelite & 1.02 & 2.35 & 1.00 & 1.93 & 0.76 & 16 \\
\midrule
\multirow{3}{*}{\textbf{DRIVE}} 
& U-Net & 24.15 & 52.15 & 1.08 & 1.88 & 0.26 & 287 \\
& U-Net Attention & 3.40 & 57.10 & 0.20 & 0.20 & 0.00 & 558 \\
& Skelite & 2.80 & 3.20 & 1.00 & 1.93 & 0.78 & 8 \\
\bottomrule
\end{tabular}
}
\label{tab:bezier-drive-comparison}
\end{table}

Table \ref{tab:bezier-drive-comparison} shows that in the Bézier dataset, our method produces thin skeletons with good topological agreement. The U-Net based approaches also exhibit few Betti errors, but prune less points producing thicker skeletons.

When transitioning to real-world data, U-Net based algorithms can produce over-thinning due to the unseen geometric conditions. This limitation is reflected in their low overlap scores and increased Betti errors. Visual analysis presented in Figure \ref{fig:drive-networks} confirms this. Conversely, Skelite demonstrates robust performance, maintaining results comparable to those on the synthetic data.

\subsection{Benchmarking Skeletonization methods} \label{sec:benchmark_skel}

\begin{figure}[t]
    \centering
    \includegraphics[width=\linewidth]{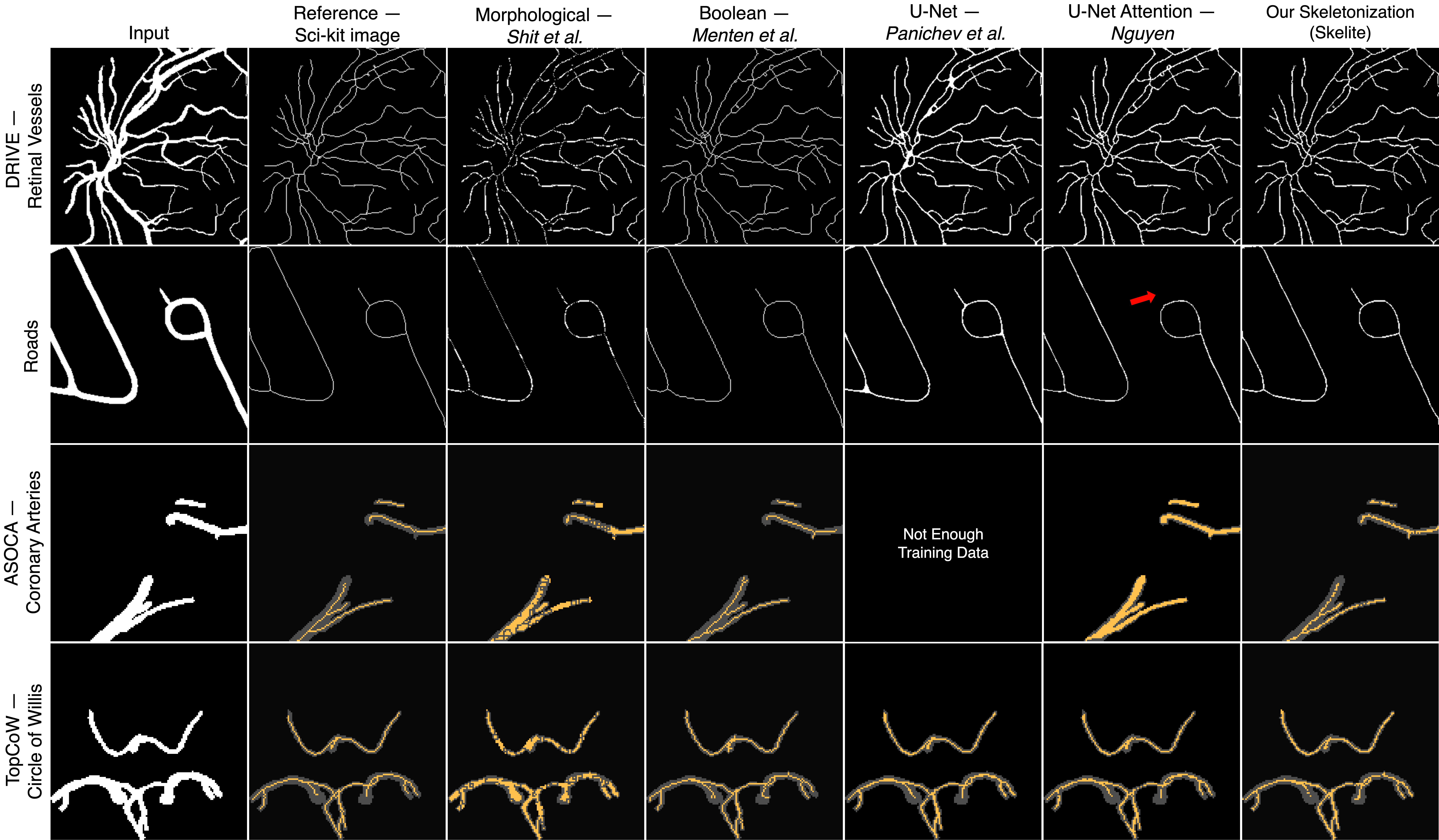}
    \caption{The result of six different skeletonization algorithms. Despite not being trained on these datasets, our method effectively produces thin skeletons in both 2D and 3D.}
    \label{fig:skeleton-benchmark-qual}
\end{figure}

We compare our skeletonization algorithm to four baselines for differentiable skeletonization: the Morphological method by Shit \textit{et al.}, the Boolean method by Menten \textit{et al.}, and the two U-Net skeletonization algorithms from Panichev \textit{et al.}~\cite{panichev_unet_skel} (denoted as U-Net) and Nguyen~\cite{nguyen_unet_skel} (denoted as U-Net Attention) trained on \textit{task-specific data}.


Results in Figure \ref{fig:skeleton-benchmark-qual} highlight the robustness of our method. Despite not being exposed to domain samples before testing, our approach generates thin skeletons that closely align with the reference across the diverse datasets. Skelite preserves topology reasonably well with only a few disconnections. This performance is comparable to U-Net based methods, which, unlike ours, have in-domain exposure and significantly more parameters. Even then, the U-Net methods tend to overfit to specific structures in the data so that they miss entire details, as highlighted by the arrow in the ROADS dataset. Furthermore, these neural network methods struggle with insufficient data in ASOCA, producing thick skeletons or failing to converge all together. Morphology-based skeletonization, on the other hand, generates many disconnected points along the medial axis. In 3D, this method exhibits the same disconnections and, in addition, fails to produce thin skeletons.

\begin{table}[b]
\centering
\caption{Quantitative comparison of topological accuracy, structural properties, and run time of skeletonization algorithms on four datasets.}
\resizebox{\textwidth}{!}{%
\setlength{\tabcolsep}{8pt} 
\begin{tabular}{@{}llcccccc@{}}
\toprule
Dataset & Method & $\beta_0$ error & $\beta_1$ error & Avg. Thickness & 99\textsuperscript{th} Max Thickness  & Dice & Run time [ms] \\
\midrule
\multirow{5}{*}{\textbf{DRIVE}} 
& Morphological & 1014.00 & 55.10 & 1.00 & 1.00 & 0.75 & 1.1 \\
& Boolean & 0.00 & 0.00 & 1.00 & 1.40 & 0.74 & 827.0 \\
& U-Net Panichev & 0.55 & 1.80 & 1.00 & 2.98 & 0.77 & 51.0 \\
& U-Net Attention & 0.50 & 1.45 & 1.00 & 2.18 & 0.78 & 400.0 \\
\cmidrule(lr){2-8}
& Ours - Skelite & 2.80 & 3.20 & 1.00 & 1.93 & 0.78 & 8 \\
\midrule
\multirow{5}{*}{\textbf{ROADS}} 
& Morphological & 686.76 & 37.69 & 1.00 & 1.22 & 0.76 & 1.2 \\
& Boolean & 0.00 & 0.00 & 1.00 & 1.00 & 0.71 & 2468.0 \\
& U-Net Panichev & 0.14 & 66.10 & 1.05 & 4.23 & 0.72 & 263.0 \\
& U-Net Attention & 5.71 & 3.71 & 1.00 & 1.70 & 0.88 & 272.0 \\
\cmidrule(lr){2-8}
& Ours - Skelite & 0.73 & 7.07 & 1.00 & 2.16 & 0.72 & 15.5 \\
\midrule
\multirow{5}{*}{\textbf{ASOCA}} 
& Morphological & 19.30 & 0.20 & 1.00 & 1.00 & 0.29 & 2.0 \\
& Boolean & 0.00 & 0.00 & 1.00 & 1.00 & 0.40 & 221.0 \\
& U-Net Panichev & \multicolumn{6}{c}{Insufficient Training Data} \\
& U-Net Attention & 0.21 & 0.05 & 1.15 & 1.60 & 0.21 & 1269.0 \\
\cmidrule(lr){2-8}
& Ours - Skelite & 4.40 & 0.20 & 1.00 & 1.00 & 0.45 & 7.0 \\
\midrule
\multirow{5}{*}{\textbf{TOPCOW}} 
& Morphological & 61.40 & 0.80 & 1.00 & 1.00 & 0.24 & 2.0 \\
& Boolean & 0.00 & 0.00 & 1.00 & 1.00 & 0.40 & 376.0 \\
& U-Net Panichev & 6.60 & 1.20 & 1.00 & 1.00 & 0.36 & 3253.0 \\
& U-Net Attention & 9.00 & 0.70 & 1.00 & 1.00 & 0.60 & 2081.0 \\
\cmidrule(lr){2-8}
& Ours - Skelite & 19.30 & 0.90 & 1.00 & 1.00 & 0.43 & 70.0 \\
\bottomrule
\end{tabular}
}
\label{tab:skel-benchmark}
\end{table}

We support our observations with quantitative data, presented in Table \ref{tab:skel-benchmark}. A key highlight is the computational efficiency of our method compared to existing approaches, particularly the Boolean and U-Net methods. While the Boolean method offers topological guarantees, it is computationally expensive. In contrast, our method computes results, on average, two orders of magnitude faster, making it significantly more practical for downstream applications.

Although our method does not explicitly enforce topological guarantees, it achieves a similar alignment to the reference skeleton as the Boolean method, as reflected by the comparable Dice scores. In contrast to the faster Morphological skeletonization, our approach produces significantly thinner and more topologically consistent skeletal representations in 3D. The thickness metrics and Betti errors further emphasize that our method generates skeletons that are both thin and largely connected across various tasks.


\subsection{Connectivity-Aware Segmentation}

Connectivity-aware and topology-aware segmentation methods have gained significant attention, leveraging approaches such as Persistent Homology \cite{stucki_segmentation,hu_betti_loss}, component graphs \cite{lux2025topograph,berger2024pitfalls}, and skeleton-based techniques \cite{shit_diff_skel,skel_recall} which relate to our method. We integrate our skeletonization network into clDice to evaluate its applicability in downstream tasks. For the 2D tasks, we used a vanilla U-Net, and in 3D, we used nnU-Net \cite{Isensee2020nnUNetAS}. We compare our results against a baseline trained solely on Dice loss and models incorporating clDice with skeletal supervision via the Morphological and Boolean methods. Additionally, we include the \textit{Skeleton Recall} loss to incorporate a recent advance that promotes connectivity preservation using non-differentiable skeletonization \cite{skel_recall}. The segmentation results are provided in Table \ref{tab:skeletonization-seg}.

Our method exhibits similar performance to the Boolean method in aiding supervision for segmentation but greatly reduces the computational overhead introduced by skeletonization. This makes our method perform at a speed comparable to the Morphological method. Furthermore, our method shows substantial enhancement of connectivity in 3D tasks, showing the lowest $\beta_0$ errors, and across 2D and 3D experiments, it demonstrates the highest clDice scores. Our results in terms of spatial accuracy, as shown by the Dice scores, mostly match the ones obtained through other skeletonization algorithms.



\begin{table}[b]
\centering
\caption{Segmentation performance with Dice loss only (Without) or using skeletal supervision.}
\resizebox{\textwidth}{!}{%
\setlength{\tabcolsep}{8pt} 
\begin{tabular}{@{}llccccc@{}}
\toprule
Dataset & Method & Dice & clDice & $\beta_0$ error & $\beta_1$ error & Epoch Time (s) \\
\midrule
\multirow{5}{*}{\textbf{DRIVE}} 
& Without & 0.77 & 0.84 & 100.00 & 13.00 & 0.09 \\
& Morphological & 0.82 & 0.86 & 45.60 & 15.00 & 0.18 \\
& Boolean & 0.83 & 0.86 & 50.00 & 13.00 & 12.00 \\
& Skel Recall & 0.63 & 0.85 & 48.30 & 14.80 & 0.14 \\
& Skelite & 0.83 & 0.86 & 53.00 & 14.00 & 0.17 \\
\midrule
\multirow{5}{*}{\textbf{ROADS}} 
& Without (Dice Loss) & 0.66 & 0.83 & 55.76 & 21.20 & 1.57 \\
& Morphological & 0.73 & 0.84 & 54.4 & 23.12 & 3.88 \\
& Boolean & 0.74 & 0.84 & 37.2 & 23.20 & 174.23 \\
& Skel Recall & 0.61 & 0.84 & 62.80 & 20.61 & 4.71 \\
& Skelite & 0.75 & 0.85 & 46.76 & 28.96 & 4.29 \\
\midrule
\multirow{5}{*}{\textbf{ASOCA}} 
& Without & 0.81 & 0.79 & 24.41 & 1.54 & 62.45 \\
& Morphological & 0.79 & 0.80 & 25.29 & 1.67 & 66.12 \\
& Boolean & 0.78 & 0.79 & 27.09 & 1.63 & 351.67 \\
& Skeleton Recall & 0.72 & 0.70 & 33.84 & 1.96 & 77.57 \\
& Skelite & 0.79 & 0.80 & 22.28 & 1.67 & 81.50 \\
\midrule
\multirow{5}{*}{\textbf{TOPCOW}} 
& Without & 0.87 & 0.93 & 1.53 & 1.18 & 57.70 \\
& Morphological & 0.87 & 0.93 & 1.49 & 1.09 & 62.90 \\
& Boolean & 0.86 & 0.93 & 1.51 & 0.91 & 322.20 \\
& Skeleton Recall & 0.83 & 0.92 & 1.10 & 1.56 & 75.20 \\
& Skelite & 0.84 & 0.95 & 1.03 & 1.36 & 77.80 \\
\bottomrule
\end{tabular}
}
\label{tab:skeletonization-seg}
\end{table}

\subsection{Ablation Studies}

\begin{table}[hpbt]
\centering
\caption{Ablation study on neighborhood loss for skeletonization learning. Adding this term encourages thinner outputs.}
\setlength{\tabcolsep}{8pt} 

\resizebox{\textwidth}{!}{%
\begin{tabular}{@{}llcccccc@{}}
\toprule
Dataset & Loss & $\beta_0$ error & $\beta_1$ error & Avg. Thickness & 99\textsuperscript{th} Max Thickness  & Dice \\
\midrule

\multirow{2}{*}{\textbf{2D Bézier}} 
& Without & 15.36 & 20.76 & 1.44 & 6.13 & 0.54 \\
& + Neighborhood Loss   & 3.57  & 11.46 & 1.11 & 3.35 & 0.69 \\
\midrule
\multirow{2}{*}{\textbf{DRIVE}} 
& Without & 4.95   & 10.15  & 1.16 & 3.54 & 0.66 \\
& + Neighborhood Loss    & 7.55   & 6.85   & 1.09 & 3.03 & 0.71 \\
\bottomrule
\end{tabular}
}
\label{tab:neigh_loss}
\end{table}

\begin{table}[b]
\centering
\caption{Ablation study on network inputs. Adding the image and skeleton as context improves the connectedness of predictions.}
\resizebox{\textwidth}{!}{%
\setlength{\tabcolsep}{8pt} 
\begin{tabular}{@{}llccccc@{}}
\toprule
Dataset & Input & $\beta_0$ error & $\beta_1$ error & Avg. Thickness & 99\textsuperscript{th} Max Thickness & Dice \\
\midrule
\multirow{3}{*}{\textbf{2D Bézier}} 
& Bound Only                      & 32.34    & 22.05   & 1.03              & 2.28          & 0.72     \\
& Bound + Image                  & 3.70    & 9.96    & 1.02              & 2.24          & 0.74     \\
& Image + Bound + Skeleton        & 2.61    & 9.48    & 1.10              & 3.10          & 0.70     \\
\midrule
\multirow{3}{*}{\textbf{Drive}} 
& Bound Only                      & 96.85   & 26.2    & 1.02              & 2.38          & 0.75     \\
& Bound + Image                   & 5.85    & 4.49    & 1.01              & 2.27          & 0.76     \\
& Bound + Image + Skeleton        & 4.20    & 4.45    & 1.07              & 2.63          & 0.72     \\
\bottomrule
\end{tabular}
}
\label{tab:net_inputs}
\end{table}

In this section, we present ablation studies to evaluate different components of the framework used to train Skelite. Table \ref{tab:neigh_loss} presents the effect of using the neighborhood loss, which encourages thinner outputs in both seen and unseen domains. Additionally, we evaluate the impact of modifying the input combinations to the CNN module of Skelite. As shown in Table \ref{tab:net_inputs}, incorporating the skeleton and image reduce disconnections in the outputs and improve topological agreement, as emphasized by the reduced Betti errors.

\section{Conclusion}

We introduce Skelite, a work that bridges the gap between neural networks and iterative skeletonization. We have shown with an extensive set of experiments that Skelite is (a) computationally efficient, (b) learns to produce thin, connected skeletons, and (c) generalizes across domains without retraining. 

Skelite abandons the path of explicit topological guarantees of the Boolean method and instead learns to produce skeletal representations from data, obtaining a significant speed-up. The obtained skeletons demonstrate good connectivity and thinness, unlike Morphology-based skeletonization, which presents frequent disconnections and poor thinning in 3D. Finally, whereas U-Net-based methods require retraining for each domain, our approach performs strongly on unseen real-world scenarios despite being trained exclusively on synthetic data. This property is a considerable advantage for domains lacking enough training data.

To the best of our knowledge, this work introduces the first learnable iterative skeletonization method. Future works could research the outcomes of using synthetic datasets emphasizing other properties such as discontinuous and jagged curves. Moreover, the proposed method could benefit from a class of data augmentations that do not introduce aliasing artifacts \cite{topology_downsample}, which can disconnect skeleton labels. Extensions to the network architecture or point proposal system could also be explored.

\subsubsection{\ackname} This research was partially supported through the research grant between Brainlab AG and Technical university of Munich. Martin J. Menten is funded by the German Research Foundation under project 532139938.




\clearpage
%
%
%
\bibliographystyle{splncs04}
\bibliography{mybibliography}

\begin{thebibliography}{10}
\providecommand{\url}[1]{\texttt{#1}}
\providecommand{\urlprefix}{URL }
\providecommand{\doi}[1]{https://doi.org/#1}

\bibitem{clCe}
Acebes, C., Moustafa, A.H., Camara, O., Galdran, A.: The centerline-cross
  entropy loss for vessel-like structure segmentation: Better topology
  consistency without sacrificing accuracy. In: International Conference on
  Medical Image Computing and Computer-Assisted Intervention (2024),
  \url{https://api.semanticscholar.org/CorpusID:273374503}

\bibitem{berger2024pitfalls}
Berger, A.H., Lux, L., Weers, A., Menten, M., Rueckert, D., Paetzold, J.C.:
  Pitfalls of topology-aware image segmentation. arXiv preprint
  arXiv:2412.14619  (2024)

\bibitem{bertrand_bool_char}
Bertrand, G.: A boolean characterization of three-dimensional simple points.
  Pattern Recognition Letters  \textbf{17}(2),  115--124 (1996).
  \doi{https://doi.org/10.1016/0167-8655(95)00100-X},
  \url{https://www.sciencedirect.com/science/article/pii/016786559500100X}

\bibitem{bloomberg_thickening}
Bloomberg, D.S.: Connectivity-preserving morphological image transformations.
  In: Other Conferences (1991),
  \url{https://api.semanticscholar.org/CorpusID:122767000}

\bibitem{blum_grassfire}
Blum, H.: A transformation for extracting new descriptors of shape (1967),
  pages: 362-380 Publication Title: Models for the Perception of Speech and
  Visual Form Place: Cambridge, MA

\bibitem{Bucila2006ModelC}
Bucila, C., Caruana, R., Niculescu-Mizil, A.: Model compression. In: Knowledge
  Discovery and Data Mining (2006),
  \url{https://api.semanticscholar.org/CorpusID:11253972}

\bibitem{topology_downsample}
Chen, C.C., Peng, C.H.: Topology-preserving downsampling of binary images.
  ArXiv  \textbf{abs/2407.17786} (2024),
  \url{https://api.semanticscholar.org/CorpusID:271431891}

\bibitem{couprie_asymmetric_simple_skel}
Couprie, M., Bertrand, G.: Asymmetric parallel 3d thinning scheme and
  algorithms based on isthmuses. Pattern Recognition Letters  \textbf{76} (04
  2015). \doi{10.1016/j.patrec.2015.03.014}

\bibitem{Skelneton}
Demir, I., Hahn, C., Leonard, K., Morin, G., Rahbani, D., Panotopoulou, A.,
  Fondevilla, A., Balashova, E., Durix, B., Kortylewski, A.: Skelneton 2019:
  Dataset and challenge on deep learning for geometric shape understanding. In:
  Proceedings of the IEEE Conference on Computer Vision and Pattern Recognition
  Workshops. pp.~0--0 (2019)

\bibitem{Fridman2004ExtractingBT}
Fridman, Y., Pizer, S.M., Aylward, S.R., Bullitt, E.: Extracting branching
  tubular object geometry via cores. Medical image analysis  \textbf{8 3},
  169--76 (2004), \url{https://api.semanticscholar.org/CorpusID:469634}

\bibitem{Asoca}
Gharleghi, R., Adikari, D., Ellenberger, K.A., Ooi, S.Y., Ellis, C., Chen,
  C.M., Gao, R., He, Y., Hussain, R., Lee, C.Y., Li, J., Ma, J., Nie, Z.,
  de~Oliveira, B.W., Qi, Y., Skandarani, Y., Wang, X., Yang, S., Sowmya, A.,
  Beier, S.: Automated segmentation of normal and diseased coronary arteries -
  the asoca challenge. Computerized medical imaging and graphics : the official
  journal of the Computerized Medical Imaging Society  \textbf{97},  102049
  (2022), \url{https://api.semanticscholar.org/CorpusID:246988905}

\bibitem{Hinton2015DistillingTK}
Hinton, G.E., Vinyals, O., Dean, J.: Distilling the knowledge in a neural
  network. ArXiv  \textbf{abs/1503.02531} (2015),
  \url{https://api.semanticscholar.org/CorpusID:7200347}

\bibitem{hu_betti_loss}
Hu, X., Li, F., Samaras, D., Chen, C.: Topology-preserving deep image
  segmentation. CoRR  \textbf{abs/1906.05404} (2019),
  \url{http://arxiv.org/abs/1906.05404}

\bibitem{Isensee2020nnUNetAS}
Isensee, F., Jaeger, P.F., Kohl, S.A.A., Petersen, J., Maier-Hein, K.: nnu-net:
  a self-configuring method for deep learning-based biomedical image
  segmentation. Nature Methods  \textbf{18},  203 -- 211 (2020),
  \url{https://api.semanticscholar.org/CorpusID:227947847}

\bibitem{image_registration}
Kim, H.C., Min, B.G., Lee, M.M., Seo, J.D., Lee, Y.W., Han, M.C.: Estimation of
  local cardiac wall deformation and regional wall stress from biplane coronary
  cineangiograms. IEEE Transactions on Biomedical Engineering
  \textbf{BME-32}(7),  503--512 (1985). \doi{10.1109/TBME.1985.325567}

\bibitem{skel_recall}
Kirchhoff, Y., Rokuss, M.R., Roy, S., Kovacs, B., Ulrich, C., Wald, T., Zenk,
  M., Vollmuth, P., Kleesiek, J., Isensee, F., Maier-Hein, K.: Skeleton recall
  loss for connectivity conserving and resource efficient segmentation of thin
  tubular structures (2024), \url{https://arxiv.org/abs/2404.03010}

\bibitem{coral_metrics}
Kruszy, K., van Liere, R., Kaandorp, J.: Quantifying differences in
  skeletonization algorithms: A case study. Proceedings of the 5th IASTED
  International Conference on Visualization, Imaging, and Image Processing,
  VIIP 2005  (01 2005)

\bibitem{systematic_evaluation_skel}
Lee, S.W., Lam, L., Suen, C.Y.: A systematic evaluation of skeletonization
  algorithms. Int. J. Pattern Recognit. Artif. Intell.  \textbf{7},  1203--1225
  (1993), \url{https://api.semanticscholar.org/CorpusID:40522276}

\bibitem{Lee1994BuildingSM}
Lee, T.C., Kashyap, R.L., Chu, C.N.: Building skeleton models via 3-d medial
  surface/axis thinning algorithms. CVGIP Graph. Model. Image Process.
  \textbf{56},  462--478 (1994),
  \url{https://api.semanticscholar.org/CorpusID:35388240}

\bibitem{Lidayov2017SkeletonbasedFF}
Lidayov{\'a}, K., Frimmel, H., Wang, C., Bengtsson, E., Smedby, {\"O}.:
  Skeleton-based fast, fully automated generation of vessel tree structure for
  clinical evaluation of blood vessel systems (2017),
  \url{https://api.semanticscholar.org/CorpusID:113501346}

\bibitem{focal_loss}
Lin, T.Y., Goyal, P., Girshick, R., He, K., Dollár, P.: Focal loss for dense
  object detection (2018), \url{https://arxiv.org/abs/1708.02002}

\bibitem{dynamic_skel_edge_detection}
Liu, J.J., Hou, Q., Cheng, M.M.: Dynamic feature integration for simultaneous
  detection of salient object, edge, and skeleton. IEEE Transactions on Image
  Processing  \textbf{29},  8652--8667 (2020). \doi{10.1109/TIP.2020.3017352}

\bibitem{Lobregt_skel}
Lobregt, S., Verbeek, P.W., Groen, F.C.A.: Three-dimensional skeletonization:
  Principle and algorithm. IEEE Transactions on Pattern Analysis and Machine
  Intelligence  \textbf{PAMI-2},  75--77 (1980),
  \url{https://api.semanticscholar.org/CorpusID:19014532}

\bibitem{lux2025topograph}
Lux, L., Berger, A.H., Weers, A., Stucki, N., Rueckert, D., Bauer, U.,
  Paetzold, J.C.: Topograph: An efficient graph-based framework for strictly
  topology preserving image segmentation. In: The Thirteenth International
  Conference on Learning Representations (2025),
  \url{https://openreview.net/forum?id=Q0zmmNNePz}

\bibitem{menten_skel}
Menten, M.J., Paetzold, J.C., Zimmer, V.A., Shit, S., Ezhov, I., Holland, R.,
  Probst, M., Schnabel, J.A., Rueckert, D.: A skeletonization algorithm for
  gradient-based optimization (2023), \url{https://arxiv.org/abs/2309.02527}

\bibitem{roads_dataset}
Mnih, V.: Machine Learning for Aerial Image Labeling. Ph.D. thesis, University
  of Toronto (2013)

\bibitem{nguyen_unet_skel}
Nguyen, N.: U-net based skeletonization and bag of tricks. pp. 2105--2109 (10
  2021). \doi{10.1109/ICCVW54120.2021.00238}

\bibitem{similarity_metric_skel}
Németh, G., Kovács, G., Fazekas, A., Palagyi, K.: A method for quantitative
  comparison of 2d skeletons  \textbf{13},  123--142 (01 2016)

\bibitem{palagyi_sequential_simple_skel}
Palagyi, K., Balogh, E., Kuba, A., Halmai, C., Erdőhelyi, B., Sorantin, E.,
  Hausegger, K.: A sequential 3d thinning algorithm and its medical
  applications. pp. 409--415 (06 2001). \doi{10.1007/3-540-45729-1_42}

\bibitem{palagyi_subiteration_simple_skel}
Pal{\'a}gyi, K., Kuba, A.: A parallel 3d 12-subiteration thinning algorithm.
  Graph. Model. Image Process.  \textbf{61},  199--221 (1999),
  \url{https://api.semanticscholar.org/CorpusID:14259397}

\bibitem{panichev_unet_skel}
Panichev, O., Voloshyna, A.: U-net based convolutional neural network for
  skeleton extraction. pp. 1186--1189 (06 2019). \doi{10.1109/CVPRW.2019.00157}

\bibitem{unet}
Ronneberger, O., Fischer, P., Brox, T.: U-net: Convolutional networks for
  biomedical image segmentation. CoRR  \textbf{abs/1505.04597} (2015),
  \url{http://arxiv.org/abs/1505.04597}

\bibitem{shit_diff_skel}
Shit, S., Paetzold, J.C., Sekuboyina, A., Zhylka, A., Ezhov, I., Unger, A.,
  Pluim, J.P.W., Tetteh, G., Menze, B.H.: cldice - a topology-preserving loss
  function for tubular structure segmentation. CoRR  \textbf{abs/2003.07311}
  (2020), \url{https://arxiv.org/abs/2003.07311}

\bibitem{Drive_Dataset}
Staal, J., Abr{\`a}moff, M.D., Niemeijer, M., Viergever, M.A., Van~Ginneken,
  B.: Ridge-based vessel segmentation in color images of the retina. IEEE
  transactions on medical imaging  \textbf{23}(4),  501--509 (2004)

\bibitem{stucki_segmentation}
Stucki, N., Paetzold, J.C., Shit, S., Menze, B., Bauer, U.: Topologically
  faithful image segmentation via induced matching of persistence barcodes
  (2022), \url{https://arxiv.org/abs/2211.15272}

\bibitem{scikit}
Van~der Walt, S., Sch{\"o}nberger, J.L., Nunez-Iglesias, J., Boulogne, F.,
  Warner, J.D., Yager, N., Gouillart, E., Yu, T.: scikit-image: image
  processing in python. PeerJ  \textbf{2}, ~e453 (2014)

\bibitem{nguyen_cbam}
Woo, S., Park, J., Lee, J.Y., Kweon, I.S.: Cbam: Convolutional block attention
  module (2018), \url{https://arxiv.org/abs/1807.06521}

\bibitem{TopCow}
Yang, K., Musio, F., Ma, Y., Juchler, N., Paetzold, J.C., Al-Maskari, R.,
  H{\"o}her, L., Li, H.B., Hamamci, I.E., Sekuboyina, A.K., Shit, S., Huang,
  H., Waldmannstetter, D., Kofler, F., Navarro, F., Menten, M.J., Ezhov, I.,
  Rueckert, D., Vos, I.N., Ruigrok, Y.M., Velthuis, B.K., Kuijf, H.J.,
  H{\"a}mmerli, J., Wurster, C., Bijlenga, P., Westphal, L.P., Bisschop, J.,
  Colombo, E., Baazaoui, H., Makmur, A., Hallinan, J., Wiestler, B., Kirschke,
  J.S., Wiest, R., Montagnon, E., L{\'e}tourneau-Guillon, L., Galdran, A.,
  Galati, F., Falcetta, D., Zuluaga, M.A., Lin, C., Zhao, H., Zhang, Z., Ra,
  S., Hwang, J., Park, H., Chen, J., Wodziński, M., M{\"u}ller, H., et~al.:
  Benchmarking the cow with the topcow challenge: Topology-aware anatomical
  segmentation of the circle of willis for cta and mra. ArXiv  (2023),
  \url{https://api.semanticscholar.org/CorpusID:267050106}

\bibitem{zhang_simple_skel}
Zhang, T.Y., Suen, C.Y.: A fast parallel algorithm for thinning digital
  patterns. Commun. ACM  \textbf{27},  236--239 (1984),
  \url{https://api.semanticscholar.org/CorpusID:39713481}

\end{thebibliography}
%




\end{document}